\documentclass[aps,prl,reprint,superscriptaddress
]{revtex4-1}
\usepackage{graphicx}
\usepackage{dcolumn}
\usepackage{rotating}
\usepackage{amssymb}
\usepackage{mathptmx}
\usepackage{amsfonts}
\usepackage{amsmath}
\usepackage{bm}
\usepackage[export]{adjustbox}
\usepackage{caption}
\usepackage{subcaption}
\usepackage{siunitx}
\usepackage{url}
\begin{document}
\title{Tuning the orbital-lattice fluctuations in the mixed spin-dimer system Ba$_{3-x}$Sr$_{x}$Cr$_{2}$O$_{8}$}

\author{Alsu Gazizulina}
\email{alsu@physik.uzh.ch}
\affiliation{Physik-Institut, Universit\"{a}t Z\"{u}rich, 8057 Z\"{u}rich, Switzerland}

\author{Diana Lucia Quintero-Castro}
\affiliation{Department of Mathematics and Physics, University of Stavanger, 4036 Stavanger, Norway}
\affiliation{Helmholtz Zentrum Berlin f\"{u}r Materialien und Energie, 14109 Berlin, Germany}

\author{Dirk Wulferding}
\affiliation{Institute for Condensed Matter Physics, TU Braunschweig, 38106 Braunschweig, Germany}
\affiliation{Laboratory for Emerging Nanometrology (LENA), TU Braunschweig, 38106 Braunschweig, Germany}

\author{Jeremie Teyssier}
\affiliation{Department of Quantum Matter Physics, University of Geneva, 1211 Geneva, Switzerland}

\author{Karel Prokes}
\affiliation{Helmholtz Zentrum Berlin f\"{u}r Materialien und Energie, 14109 Berlin, Germany}

\author{Fabiano Yokaichiya}
\affiliation{Helmholtz Zentrum Berlin f\"{u}r Materialien und Energie, 14109 Berlin, Germany}

%\author{Dmitry Chernyshov}
%\affiliation{Swiss-Norwegian Beamline, European Synchrotron Radiation Facility, 38043 Grenoble, France}

\author{Andreas Schilling}
\affiliation{Physik-Institut, Universit\"{a}t Z\"{u}rich, 8057 Z\"{u}rich, Switzerland}

%\date{\today}

\begin{abstract}
In $A_{3}$Cr$_{2}$O$_{8}$, where $A$ = Sr or Ba, the Cr$^{5+}$ ions surrounded by oxygen ions in a tetrahedral coordination are Jahn-Teller active. The Jahn-Teller distortion leads to a structural transition and a related emergence of three twinned monoclinic domains below the structural phase transition. This transition is highly \textit{dynamic} over an extended temperature range for $A$ = Sr. 
%The onset of orbital ordering results directly from the combination of exchange correlation within the $3d^{1}$ shell of Cr$^{5+}$ with the monoclinic distortion. 
We have investigated mixed compounds Ba$_{3-x}$Sr$_{x}$Cr$_{2}$O$_{8}$ with $x=2.9$ and $x=2.8$ by means of X-ray and neutron diffraction, Raman scattering and calorimetry. Based on the obtained evolution of the phonon frequencies, we find a distinct suppression of the orbital-lattice fluctuation regime with increasing Ba content. This stands in contrast to the linear behaviour exhibited by unit cell volumes, atomic positions and intradimer spin-spin exchange interactions. 

\end{abstract}

\maketitle	

\section{\label{Intr}Introduction}
%General interest: 
%Some gapped quantum systems exhibit a magnetic phase transition which strongly related to the environment of the magnetic ions~\cite{zapf}. This magnetic phase transition is a result of  electronic correlations. The orbital-lattice coupling in $3d$ orbitals has been intensively investigated. The $3d$ orbital systems can be classified by the electronic structures of the ground state. 
In many condensed matter systems, e.g. manganites, ferrites, cuprates, among others, the key to functional magneto-electronic properties is the interaction between degenerate orbital degrees of freedom and lattice fluctuations through a cooperative Jahn-Teller effect~\cite{Varignon_perovskite, Stroppa_adv, Kyono_spinels, tokura_nagaosa, keller, lee}.  This effect can eventually change the crystallographic structure, fully lift the orbital degeneracy or create a short-range orbital order through a dynamical Jahn-Teller distortion. Spin-spin correlations are also relevant in these strongly correlated compounds, and can interact with orbital and lattice degrees of freedom. Even more, interesting dynamics can emerge when the spin-spin correlations are spatially frustrated, leading to a macroscopic degeneracy of the ground state \cite{Santanu_spinel}. \par

\begin{figure}
\includegraphics[width=0.5\textwidth]{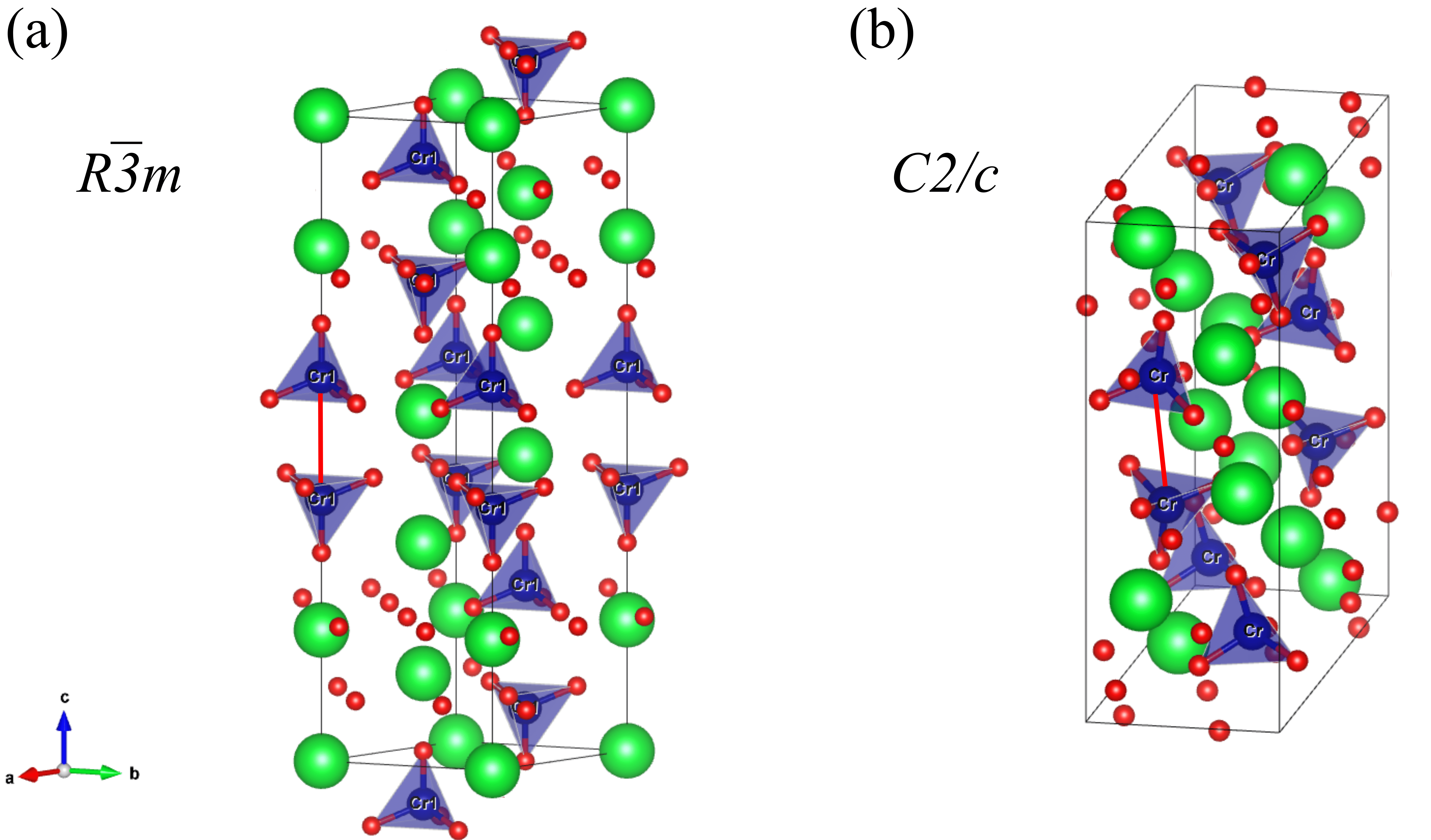}
\caption{(Color online) (a) Hexagonal and (b) monoclinic unit cells in $A_{3}$Cr$_{2}$O$_{8}$. Cr$^{5+}$ ions, represented in blue, are located in oxygen tetrahedra, O$^{2-}$ ions are represented by red spheres and $A$ = Sr/Ba by green spheres. Dimer bonds are represented in red}
\label{structure_both}
\end{figure}

%Specific interest: 
All these competing degrees of freedom are present in $A_{3}$Cr$_{2}$O$_{8}$ (with $A$ = Ba, Sr). These systems consist of a 3-dimensional network of coupled spin dimers formed by the Jahn-Teller active ion, Cr$^{5+}$. The dimers form hexagonal bilayers and are antiferromagnetically linked in a triangular in-plane arrangement, resulting in a certain degree of magnetic frustration. 
In an earlier work, Raman scattering experiments were performed on a single crystal of Sr$_{3}$Cr$_{2}$O$_{8}$~\cite{dirk}. 
An extended orbital-lattice fluctuation regime was detected and related to structural distortion, orbital ordering and further changes in the orbital sector below the Jahn-Teller transition temperature $T_{JT}$. At this temperature,  $T_{JT}$~=~285~K for Sr$_{3}$Cr$_{2}$O$_{8}$ and $T_{JT}$~=~70~K for Ba$_{3}$Cr$_{2}$O$_{8}$, the structural transition from hexagonal $R\bar{3}m$ to monoclinic $C2/c$ symmetry appears in both compounds (Fig.~\ref{structure_both}) ~\cite{chapon}. 
There is no fluctuation regime in the lattice degrees of freedom in Ba$_{3}$Cr$_{2}$O$_{8}$~\cite{wang-12} as the orbital entropy associated with the splitting of the orbital ground state is completely recovered at the Jahn-Teller transition. These differences in the dynamic orbital properties between these two systems stand in contrast to the magnetic ground state (singlet) and magnetic excitations (triplons) which are essentially similar in both compounds~\cite{kofu, diana_mag}. %Therefore, the orbital entropy associated with the splitting of the orbital ground state is completely recovered at the Jahn-Teller transition.

% Mixed systems
It is known that by mixing different cations in the atomic structure of functional materials, weak disorder can be introduced and it may be sufficient to change physical properties dramatically~\cite{tokura_nagaosa, dirty}.  Many exotic phases can be observed in mixed condensed matter electronic systems, such as perovskites~\cite{phase_maezono} and cuprates~\cite{keller}, where unusual strong orbit-lattice coupling has been observed~\cite{goodenough_perovskite}.  The abundance of functional properties in those mixed systems is stimulated by a static displacive disorder due to differences in ionic radii, which creates mixed charge distribution~\cite{dagotto}. However, the complex magnetic phase diagrams by settling of very different long range magnetic order as a function of cation mixing present in many of these compounds, make the direct study of the orbital dynamics non-trivial.

%Here, we report \textit{dynamic} structural phase distortions in the mixed system Ba$_{3-x}$Sr$_{x}$Cr$_{2}$O$_{8}$. It is known that in mixed systems, even a weak disorder may be sufficient to change properties dramatically. Many exotic phases can be observed in mixed strongly correlated electronic systems, such as perovskites~\cite{phase_maezono}, where an unusually strong orbit-lattice coupling was observed~\cite{goodenough_perovskite}. The complexity in those systems is induced by an interplay of many degrees of freedom, such as charge and spacial distributions, that are sensitive to the introduced disorder~\cite{dagotto}, causing incoherent dynamics~\cite{tokura_nagaosa} and settling of different ground state. The introduction of impurities is an important topic due to the fact that quantum disordered spin systems behave differently from classical. Many important results have already been obtained in quantum disordered systems~\cite{dirty}.

%what we have done so far: 
Recently, we reported on the effect of chemical substitution on the magnetic properties of Ba$_{3-x}$Sr$_{x}$Cr$_{2}$O$_{8}$ upon partially replacing Sr by Ba by 3.3\% and 6.6\%. It was found that the intradimer spin-spin interaction is smaller in the mixed compound with $x=2.9$ than in pure Sr$_{3}$Cr$_{2}$O$_{8}$ by 4\%, whereas the interdimer exchange interactions are reduced by 7\%~\cite{Alsu_article}. % more abruptly.  
No sign of magnetic disorder was found. %Henrik's papers
Furthermore, we have found that the critical magnetic field related to the condensation of the triplet magnetic excitations decreases monotonically with $x$~\cite{grundmann_tuning}.
This change is in accordance with the change the intra-dimer interaction constant $J_0$~\cite{grundmann_structure}.

%what we do in this paper: 
Here, we report the study of the \textit{dynamic} structural phase distortions in the mixed systems Ba$_{3-x}$Sr$_{x}$Cr$_{2}$O$_{8}$ (with $x=2.8,$ $2.9$). Both pristine and mixed compounds preserve the magnetic ground state and are therefore the ideal model systems to study uniquely orbit-lattice fluctuations as a function of cation mixing.
We investigated the structural phase transition in the mixed compounds over an extended temperature range. Using heat capacity measurements, Raman scattering, neutron single crystal diffraction and high-resolution X-ray powder diffraction techniques, we have explored the variation of the crystal structure of Ba$_{3-x}$Sr$_{x}$Cr$_{2}$O$_{8}$ (with $x=2.9$ and $x=2.8$) as a function of temperature. \par

\section{\label{sec:one}Experimental details}
Single crystals of Ba$_{3-x}$Sr$_{x}$Cr$_{2}$O$_{8}$ were grown with $x=2.9$ and $x=2.8$, as described in Ref.~\cite{Alsu_article}.
Heat capacity measurements were performed in a Physical Property Measurement System (PPMS, Quantum Design Inc.).  \par

The temperature dependent high-resolution powder X-ray diffraction experiments were carried out at the Swiss-Norwegian beamline BM01 at the ESRF in Grenoble~\cite{snbl}. Temperature control was achieved using a liquid nitrogen cryostream. A powder sample of Ba$_{0.1}$Sr$_{2.9}$Cr$_{2}$O$_{8}$ was measured in the temperature range from 100~K to 300~K. To extract crystallographic lattice parameters and atomic displacement positions, Rietveld refinement was performed using the FullProf software~\cite{full}.\par

The nuclear structure of the spin dimer system Ba$_{3-x}$Sr$_{x}$Cr$_{2}$O$_{8}$ with $x=2.9$ was studied by performing single crystal neutron diffraction on the 2-axis-diffractometer E4 at the Helmholtz Zentrum f\"{u}r Materialien und Energie in Berlin. The sample with volume $\cong$ 1~cm$^3$ was mounted with ($h,h,l$)$_{h}$ in the scattering plane (subindices \textit{h} and \textit{m} correspond to hexagonal and monoclinic notation, respectively). The incident wavelength was \SI{2.4}{\angstrom}. Measurements of intensities of key Bragg peaks were made at temperatures between 1.5~K and 290~K.\par

Raman scattering experiments were performed using an excitation wavelength of $\lambda=514.5$~nm on single crystals of Ba$_{3-x}$Sr$_{x}$Cr$_{2}$O$_{8}$ with $x=2.9$ and $x=2.8$. Samples were embedded in a silver matrix to achieve better thermalization. Both crystals were mounted and polished parallel to the \textit{ac} plane. The incoming laser polarization was along the \textit{c} axis and scattered light of all polarizations was detected. The laser power was set to 0.5~mW with a spot diameter of about \SI{2}{\micro\metre}. Scans were recorded in 10~K steps from 5~K up to 200~K, followed by 20~K steps in the range 200~K to 340~K.

\section{\label{sec:res}Results and discussion}
\begin{figure}
\includegraphics[width=0.5\textwidth]{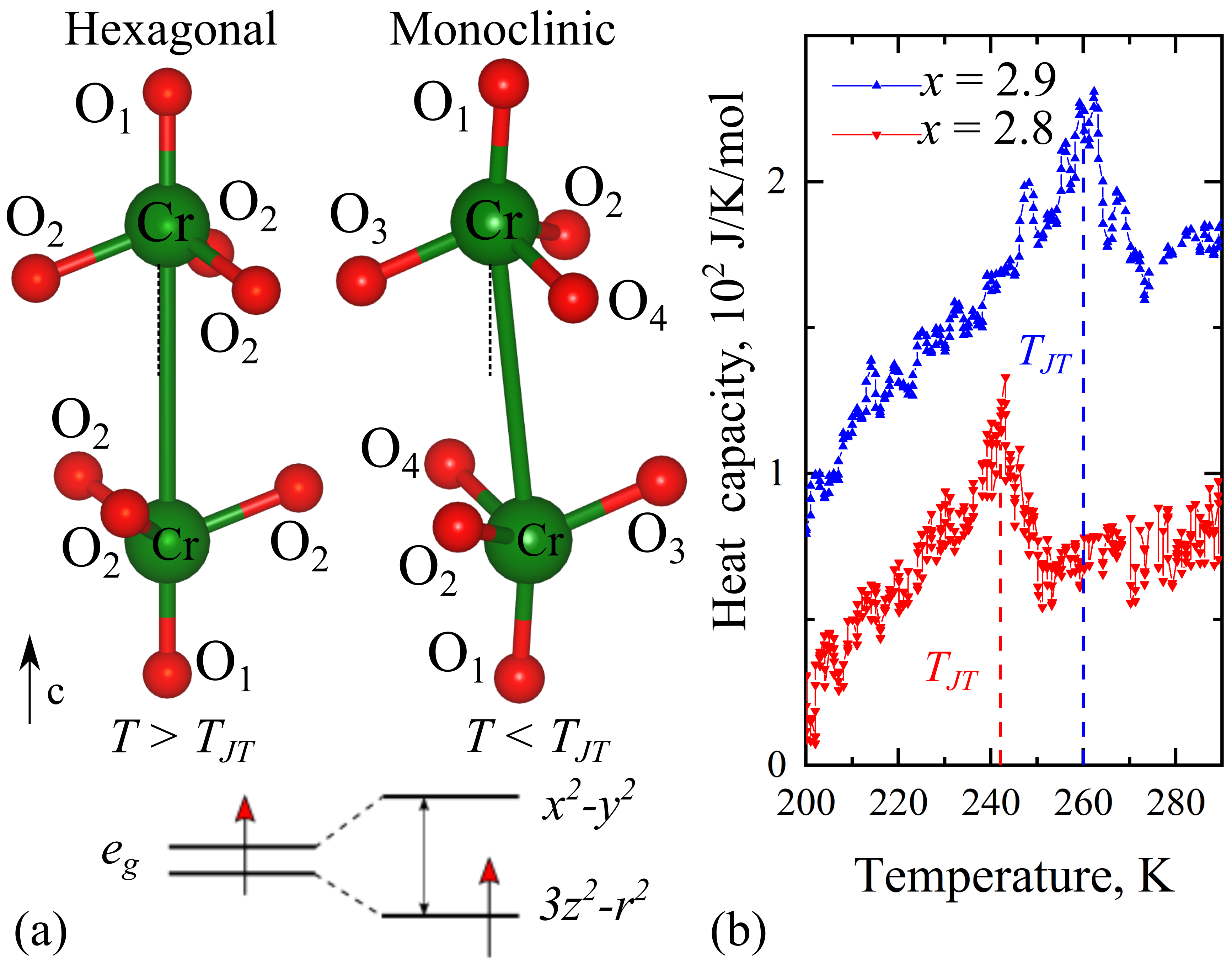}
\caption{(Color online) (a) Two CrO$_4$ tetrahedra forming a dimer unit in the hexagonal and the monoclinic phases (top). Corresponding occupation of electronic levels in each phase (bottom). (b) Heat capacity-temperature dependences for both samples: $x=2.9$ (black curve) and $x=2.8$ (red curve) showing $\lambda$-like JT-transitions at 260~K and 242~K, respectively.}
\label{hc}
\end{figure}

In the mixed compounds Ba$_{3-x}$Sr$_{x}$Cr$_{2}$O$_{8}$, the magnetic ion Cr$^{5+}$ is surrounded by four oxygen ions in a tetrahedral symmetry, as shown in Fig.~\ref{hc}(a). A cooperative Jahn-Teller distortion lifts the electronic degeneracy of tetrahedrally coordinated Cr$^{5+}$ ions. At $T > T_{JT}$, the structure is hexagonal in the space group $R\bar{3}m$. Bilayers of CrO$_4$ tetrahedra form a frustrated arrangement of two degenerate $3d^1$ $e_g$ levels ($3z^2-r^2$ and $x^2-y^2$). Each orbital is coupled to the displacement of the O atoms surrounding the ion.
For $T < T_{JT}$, crystallographic distortions lead to the monoclinic structure $C2/c$ and a splitting of the levels of the lower-lying orbitals. This structural transition is characterised by an antiferro-distortive displacement of the apical O$_1$ oxygen ions and a slight rotation of the O$_2$ ions in the tetrahedral bottom plane within each dimer. It leads to three twinned monoclinic domains in the parent compounds Sr$_{3}$Cr$_{2}$O$_{8}$~\cite{diana_mag} and Ba$_{3}$Cr$_{2}$O$_{8}$~\cite{kofu}.\par

Recently, we showed that introducing Ba in the Sr$_{3}$Cr$_{2}$O$_{8}$ system does neither lead to magnetic disorder nor to intra-gap intensities in the dispersion relation~\cite{Alsu_article}. 
%heat capacity:
The Jahn-Teller distortion temperature in Ba$_{3-x}$Sr$_{x}$Cr$_{2}$O$_{8}$ is gradually suppressed with decreasing $x$, as shown in Ref.~\cite{grundmann_influence}, and vanishes for intermediate stoichiometries $x$. This transition is reflected as a $\lambda$-like transition in the heat capacity, as measured for $x=2.9$ and $x=2.8$ [Fig.~\ref{hc}(b)]. For the mixed compounds $x=2.9$ and $x=2.8$, the Jahn-Teller transition temperatures are $T_{JT}$~=~260~K and 242~K, whereas for the pure 
Sr$_{3}$Cr$_{2}$O$_{8}$ $T_{JT}$~ is ~285~K and ~70~K for Ba$_{3}$Cr$_{2}$O$_{8}$, respectively~\cite{dirk, kofu}. 
%A correlation between structural and magnetic properties mainly determines the feature of the whole family of spin-dimerized antiferromagnets.
\par

\begin{figure}
\includegraphics[width=0.4\textwidth]{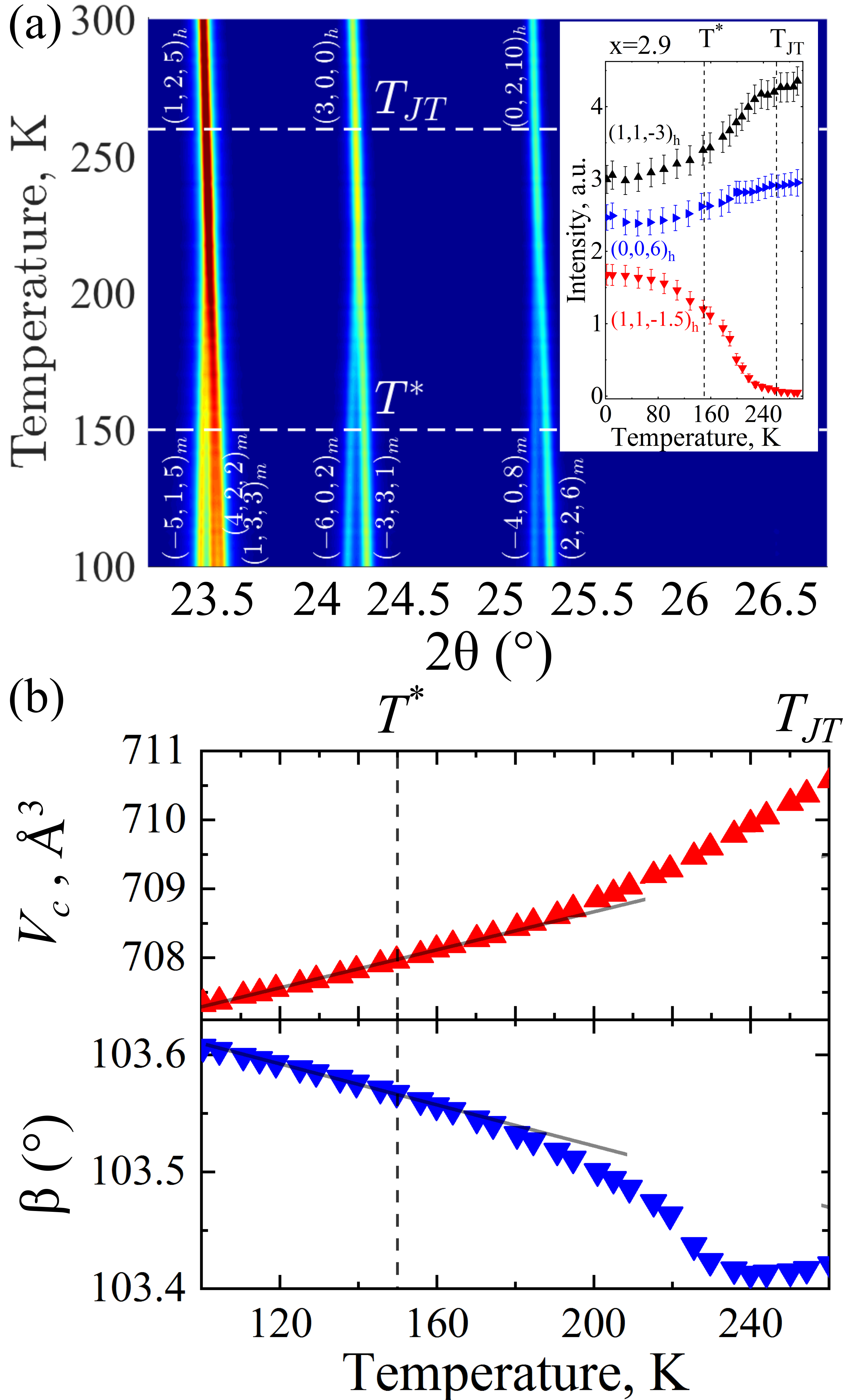}
\caption{(Color online) (a) Colour contour plot of the temperature dependence of the X-ray diffraction  pattern for $x=2.9$ powder. The centred reflection (3,0,0)$_{h}$ splits into (-6,0,2)$_{m}$ and (-3,3,1)$_{m}$ at lower temperatures.  Inset: Temperature dependence of the integrated intensity of (1,1,-3)$_{h}$, (1,1,-1.5)$_{h}$ and (0,0,6)$_{h}$ reflections of a $x=2.9$ single crystal. Dotted lines show $T_{JT}$~=~260~K and $T^{*}$~=~150~K. (b) The unit cell volume and monoclinic angle $\beta$ in the monoclinic symmetry $C2/c$ as a function of temperature (red and blue data points, respectively).  The lines are guides to the eye}
\label{E4_Ba}
\end{figure}

%Diffraction
In the X-ray powder diffraction patterns, obtained at temperatures above $T_{JT}$ for $x=2.9$ compound, the observed Bragg reflections were indexed using the space group $R\bar{3}m$. Our analysis of those patterns shows that the lattice at $T < T_{JT}$ is then better described using the space group $C2/c$, however, with new diffraction peaks gradually appearing as can be seen in Fig.~\ref{E4_Ba}(a). Our results are in agreement with density functional theory calculations of the orbital ordering in Sr$_{3}$Cr$_{2}$O$_{8}$ that demonstrate the strong electron correlation which arises within the $3d^1$ shell and can clearly explain a phase transition leading to the stabilisation of its monoclinic $C2/c$ space-group symmetry and spin-singlet magnetic ground state~\cite{radtke}.\par

The crystallographic parameters extracted by using Rietveld refinement~\cite{full} are presented in Table~\ref{mon_hex}. The weighted profile $R$-factor shows a relatively good fit.
\begin{table*}
\caption{Crystallographic parameters extracted from the Rietveld refinement of the BM01 data for $x=2.9$. The weighted profile $R$-factor is $R_{wp}\cong 8$ for both refinements, at $T < T_{JT}$ and at $T > T_{JT}$.}
\begin{ruledtabular}
\label{mon_hex}
\begin{tabular}{ccccccc}
        $x=2.9$          &  $C2/c$ (100~K)&  & $R\bar{3}m$ (300~K)   & \\ \hline
\multicolumn{1}{c}{$a$ [\AA]} & 9.60(8)  & & 5.57(7) & \\
\multicolumn{1}{c}{$b$ [\AA]}  & 5.51(6)  & & 5.57(7)  &\\ 
\multicolumn{1}{c}{$c$ [\AA]}  & 13.72(7)  & & 20.20(6)  &\\ 
\multicolumn{1}{c}{$\alpha$ [deg]} & 90 & & 90  &\\ 
\multicolumn{1}{c}{$\beta$ [deg]} &$\cong$ 103.61(3)  & &  90 &  \\ 
\multicolumn{1}{c}{$\gamma$ [deg]} & 90 & &  120 & \\ \hline
\multicolumn{1}{c}{Atom} & Atomic position [x, y, z]$_{m}$ & B$_{iso\ m}$ & Atomic position [x, y, z]$_{h}$ &B$_{iso\ h}$ \\  \hline

\multicolumn{1}{c}{Cr} & [0.2028, 0.2557, 0.8576] & 0.84(7)&  [0, 0, 0.4052]&0.79(9) \\ 
\multicolumn{1}{c}{Sr$_1$/Ba$_1$} & [0, 0.2715, 0.2500] & 0.46(1) &  [0, 0, 0] & 0.79(9)\\ 
\multicolumn{1}{c}{Sr$_2$/Ba$_2$} & [0.1020, 0.2482, 0.5550] & 0.42(8) &  [0, 0, 0.2037]  & 0.64(9)\\ 
\multicolumn{1}{c}{O$_1$} & [0.1602, 0.3086, 0.7386] & 0.38(5) &  [0, 0, 0.3255]  & 0.97(8)\\ 
\multicolumn{1}{c}{O$_2$} & [0.8624, -0.0091, 0.6078] & 0.42(4) &  [0.8353, 0.1647, 0.8981]  & 0.97(8)\\ 
\multicolumn{1}{c}{O$_3$} & [0.1143, 0.7842, 0.5987] & 0.85(6) &   \\ 
\multicolumn{1}{c}{O$_4$} & [0.3683, 0.9842, 0.5833] & 0.61(2) &   \\

%\multicolumn{1}{c}{R$_{wp}$} & $\cong$ 8 & $\cong$ 8  \\ 
\end{tabular}
\end{ruledtabular}
\end{table*}
The lattice parameters and unit cell volume shrink in a non-linear manner as a function of temperature (Figure~\ref{E4_Ba}(b)). Apart from the unit cell volume and lattice constants, the structural distortion also affects the monoclinic angle $\beta$. The monoclinic angle decreases non-linearly by $\Delta\beta\approx$~0.12$^{\circ}$ when increasing the temperature from 80~K to 220~K, as can be seen in Fig.~\ref{E4_Ba}(b), and deviates slightly from the value that is expected from a conversion between the rhombohedral and the monoclinic space groups~\cite{grundmann_structure}.
\par
The structural distortion is characterized by an antiferro-distortive displacement of the apical O$_{1}$ oxygen ion within the CrO$_4$ tetrahedra similar to Sr$_{3}$Cr$_{2}$O$_{8}$~\cite{chapon}. The displacement of O$_{1}$ is coupled to the displacement of the Sr$_{1}$/Ba$_{1}$ and a slight rotation of the tetrahedral basal plane, spanned by the O$_{2}$ ions. This rotation and distortion further modifies the Cr-O distances below $T_{JT}$ to O$_{2}$, O$_{3}$ and O$_{4}$. A modification of the Cr-O distances leads to a change in energy of the electronic orbitals of Cr$^{5+}$. The crystalline electric field splits the 3d orbitals of the Cr$^{5+}$ ion into lower-lying non-bonding $e_{g}$ orbitals and higher-lying anti-bonding $t_{2g}$ orbitals. This distortion reduces the point group symmetry of the Cr$^{5+}$ site from $C_{3v}$ (in Schoenflies notation) in the hexagonal structure to $C_1$ in the monoclinic structure where the $e_{g}$ state degeneracy is lifted, with a separation into ($3z^2-r^2$) and ($x^2-y^2$) orbitals. The oxygen tetrahedron around the Cr$^{5+}$ ion is only slightly affected by varying the temperature and changes only with the onset of the structural transition at $T_{JT}$.\par

Single crystal neutron diffraction was also performed in order to obtain detailed information about the formation of three monoclinic twins by measuring key reflections in the ($h,h,l$)$_{h}$ plane, which correspond to a unique, double or triple twin, as a function of temperature. The results are shown as an inset in Figure~\ref{E4_Ba}(a). Three reflections were measured: (1,1,-3)$_{h}$, (1,1,-1.5)$_{h}$ and (0,0,6)$_{h}$. The reflection (1,1,-1.5)$_{h}$ corresponds to two different monoclinic reflections (1,1,-1.5)$_{h}$ = (0,2,1)$_{m^{(1, 2)}}$ = (3,1,2)$_{m^{(3)}}$, where the $m^{(n)}$ notation corresponds to $n^{th}$ monoclinic twin.
%, and it is forbidden in the hexagonal symmetry, 
The reflection (0,0,6)$_{h}$ corresponds to three monoclinic twins (0,0,6)$_{h}$ = (0,0,4)$_{m^{(1, 2, 3)}}$ and it is an allowed hexagonal reflection, and the (1,1,-3)$_{h}$ reflection is allowed in both hexagonal and monoclinic and corresponds to the reflections (1,1,-3)$_{h}$ = (0,2,2)$_{m^{(1)}}$ = (3,1,3)$_{m^{(2, 3)}}$. Conversions from hexagonal to monoclinic Miller indices have been done following following Ref.~\cite{diana_mag}. 
%Therefore it should evidence a more complex temperature dependence than the other two reflections. 
The largest changes of the diffraction patterns are observed for temperatures far below the Jahn-Teller transition temperature $T_{JT}$ reaching full intensities well below $T_{JT}$.  The observed dependences indicate that the structure stabilizes below some characteristic temperature $T^{*} < T_{JT}$, which we define more precisely further below.

\begin{figure}[htt]
\includegraphics[width=0.45\textwidth]{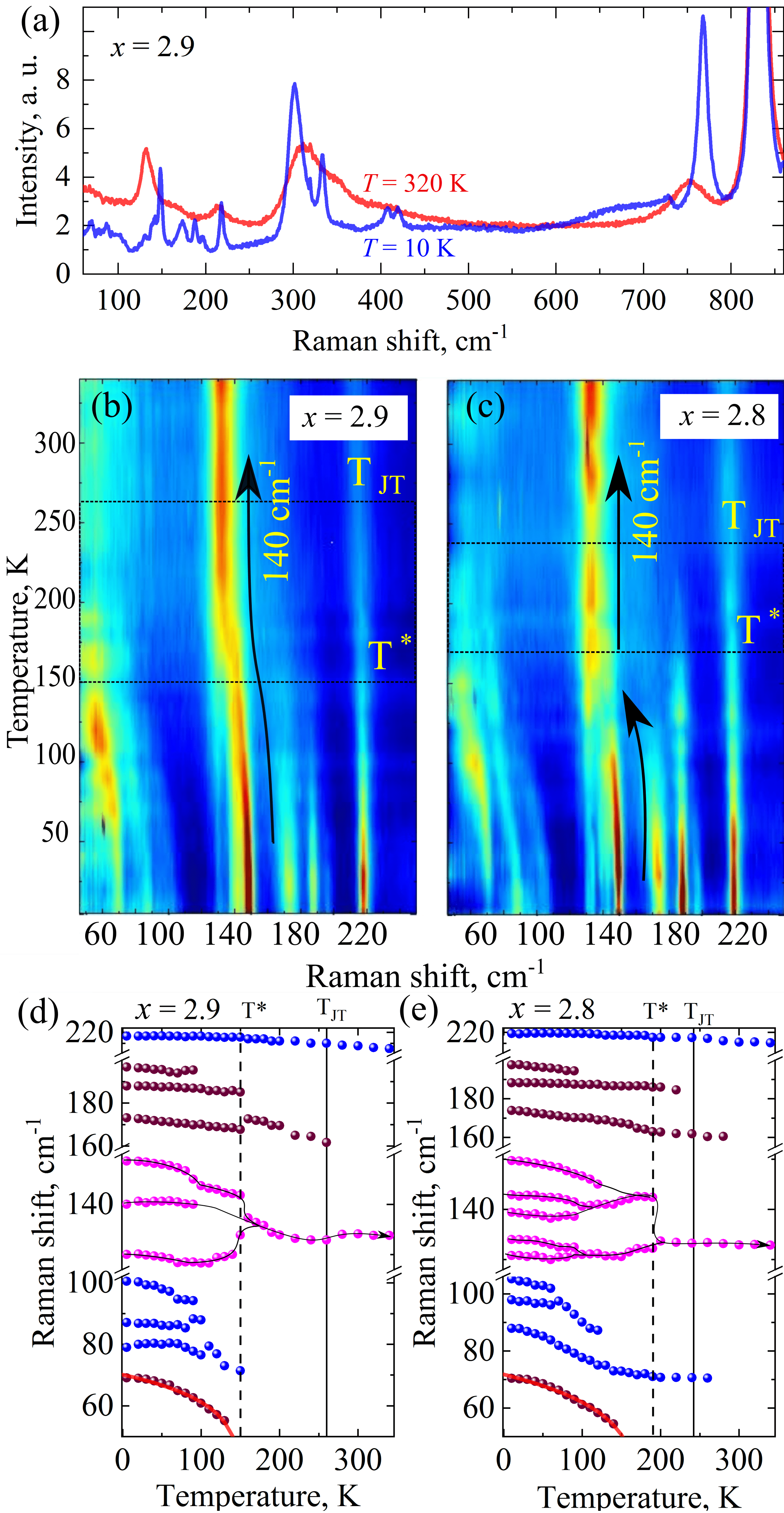}
\caption{(Color online) (a) Raman spectra collected at $T$~=~320~K (red curve) and at 10~K (blue curve) for $x=2.9$. Colour plot spectra in the temperature range from 5~K to 340~K for (b) $x=2.9$ and (c) $x=2.8$. Temperature evolution of the phonon frequencies for (d) $x=2.9$ and (e) $x=2.8$. The red line is a fit of the mode around 70 cm$^{-1}$.}
\label{2D}
\end{figure}

%Raman
%For a detailed analysis of the fluctuation regime and a comparison of samples with different Ba content
%Our Raman results for $x=2.9$ and $x=2.8$. 
Fig.~\ref{2D}(a) shows two Raman spectra collected at 310~K and 10~K. At low temperatures two modes develop between 310 and 340~cm$^{-1}$. They are rather broad as compared to other peaks. This indicates a possible mixing of the orbital excitation and a phonon. A factor group analysis~\cite{dirk} gives the number of Raman active phonons that was analysed for the parent compound  Sr$_{3}$Cr$_{2}$O$_{8}$. 11~Raman active phonon modes are expected in the hexagonal symmetry $R\bar{3}m$ and 39 corresponding modes in the monoclinic space group C2/$c$. The respective mode frequencies and their symmetry assignments are given in Ref.~[\onlinecite{dirk}]. We detected only slight shifts in the phonon frequencies when compared with the parent compound Sr$_{3}$Cr$_{2}$O$_{8}$.\par

 We observe distinct changes in the phonon spectra in the temperature range 100~K~-~260~K [see Figs.~\ref{2D}(b) and \ref{2D}(c)]. Below a temperature around 150~K ($T^{*}$) for $x=2.9$, the quasi-elastic scattering (seen as a subtle increase in scattering intensity towards low energies) is suppressed and the Raman spectra show a behaviour that can be described as a stabilisation of the lattice. Therefore, $T^{*}$ marks a crossover from orbital fluctuations to long-range and static structural distortions and we can therefore interpret the temperature window between $T^{*}$ and $T_{JT}$ as a range with strong fluctuations.
 The strong electron correlation arising within the $3d$ shell of Cr$^{5+}$ can explain the phase transition leading to the stabilisation of the monoclinic $C2/c$ space group symmetry.\par

%Now, let us focus on the distinction between the two compounds with different Ba substitutions. 
Taking a closer look at the phonon mode around 140~cm$^{-1}$, the transition is very smooth between 200~K and 150~K in the case of $x=2.9$, whereas for $x=2.8$ there seems to be a sudden jump around 125~K [see marks in Figs.~\ref{2D}(b) and \ref{2D}(c)]. We can conclude that there is a slightly different fluctuating behaviour for the lattice of two compounds.
In Fig.~\ref{2D}(d, e), the frequencies of the phonon modes, extracted with a Lorentz fitting, are plotted as a function of temperature. Both samples, $x=2.9$ and $x=2.8$, show similar trends in their temperature evolution with a strong softening of the 3A$_{g}$ modes at about 70, 90 and 105~cm$^{-1}$. In particular, the mode around 70 cm$^{-1}$ evidences a dramatic softening upon increasing the temperature towards $T^{*}$ due to a strong coupling to electronic degrees of freedom. This mode is very susceptible to the atomic displacement of the apical oxygen O$_1$ in the CrO$_4$-tetrahedra, and the related displacement may correspond to a possible rotational motion of the CrO$_4$-tetrahedra that modifies all Cr-O distances. Therefore, once the Jahn-Teller distortion changes the transition from dynamic to static at $T^{*}$, this soft mode starts to emerge. This process can be viewed as a static tetrahedral distortion. Thus, we can consider this phonon energy as a secondary order parameter, that emerges from orbital order and can be described by mean-field theory, $\omega(T)$ = A $|T^{*} - T|^{\eta}$ (see fits in Fig.~\ref{2D}(d, e)), where $\eta$ denotes the critical exponent of the secondary order parameter and $T^{*}$ indicates the critical temperature of the transition~\cite{landau}. We obtain $\eta$ = 0.12 and 0.16 for $x=2.9$ and $x=2.8$, respectively. These exponents are close to the two-dimensional Ising solution where $\eta$ = 1/8. A similar mean-field exponent has been extracted from a soft phonon mode in the related Jahn-Teller active BaNa$_2$Fe[VO$_4$]$_2$~\cite{dirk_screw}. \par

\begin{figure}[htt]
\includegraphics[width=0.5\textwidth]{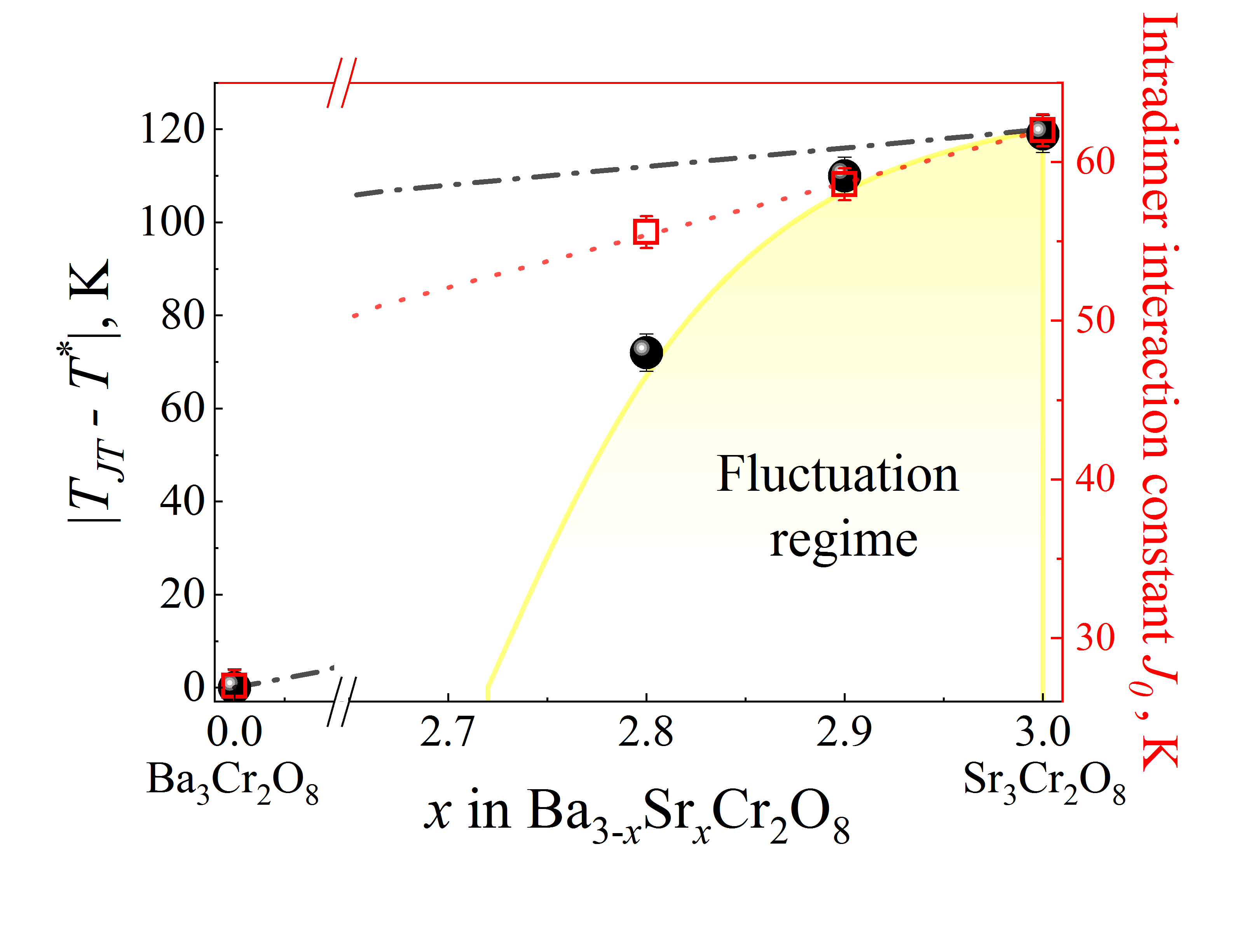}
\caption{(Color online) Temperature range of the fluctuation regime $|T _{JT}$ - $T^{*}|$ (filled black spheres) and intra-dimer interaction $J_0$ (open red squares, Ref.~[\onlinecite{Alsu_article}]) versus Sr content $x$ in Ba$_{3-x}$Sr$_{x}$Cr$_{2}$O$_{8}$. Lines in the respective color indicate a linear dependence of these quantities on the Sr content $x$ for comparison.}
\label{phase_JT}
\end{figure}

Based on the mean-field fit to the soft mode energy 70~cm$^{-1}$, we obtain $T^{*}$ = (150~$\pm$~3)~K for $x=2.9$ and $T^{*}$ = (170~$\pm$~5)~K for $x=2.8$. Interestingly, $T^{*}$ for $x=2.8$ is noticeably enhanced in comparison with $x=2.9$. This means that the fluctuation regime is narrowed with increasing Ba content and fluctuations are suppressed. Fig.~\ref{phase_JT} shows the temperature range of the fluctuation regime $|T _{JT}$ - $T^{*}|$ and the intra-dimer magnetic exchange interaction $J_0$ as a function of Sr content $x$ in Ba$_{3-x}$Sr$_{x}$Cr$_{2}$O$_{8}$. Values for the end members Ba$_3$Cr$_2$O$_8$ and Sr$_3$Cr$_2$O$_8$ are taken from literature~\cite{wang-12, dirk}. The interaction $J_0$ changes monotonously in the shown region of $x$ but not linearly as the lattice parameters~\cite{grundmann_structure}. The area below the data points of the temperature range $|T _{JT}$ - $T^{*}|$ marked by yellow color is a guide for the eye.
It shows that the fluctuation regime decreases drastically and would vanish around $x=2.72$.
\par

There are different plausible scenarios leading to a decrease in $T_{JT}$ and a suppression of the fluctuation regime upon increasing Ba content: One possibility could be some disorder or bond length mixing induced by Ba/Sr substitution. However, our experiments evidence a systematic trend, where the end members (with the lowest level of disorder) mark the two extreme cases of no fluctuations vs. maximum fluctuations. Since lattice constants and unit cell volume change linearly~\cite{grundmann_structure} as a function of $x$, there is also no obvious structural reason for the suppression of $T^{*}$. A more likely scenario is a continuous tuning of the intradimer interaction $J_0$ (shown on the right axis of Fig.~\ref{phase_JT}) as well as the ratio of inter- to intradimer interaction $J'/J_0$ with $x$. In fact it was found that for Ba$_3$Cr$_2$O$_8$ this ratio is slightly larger ($J'/J_0$ = 0.82) than for Sr$_3$Cr$_2$O$_8$ ($J'/J_0$ = 0.64)~\cite{Alsu_article}. This subtle difference could be the reason for the fundamentally different dynamics. An observed stronger interdimer coupling $J'$ (with respect to $J_0$) can lead to a faster stabilization of the spin- and the orbital subsystem, whereas a weaker interdimer coupling preserves a low-dimensional quantum magnetism character of the spin dimer system, which is more prone to strong fluctuations.

Lattice dynamics can be present in the hexagonal lattice in order to achieve a thermodynamically favoured orbitally nondegenerate ground state of a Jahn-Teller active ion in a distorted environment.  Similar to Ba$_{3-x}$Sr$_{x}$Cr$_{2}$O$_{8}$, a dynamical transition has been observed in different materials, such as BaNa$_2$Fe[VO$_4$]$_2$ where the transition is induced by the gradual evolution of the distortion around FeO$_6$~\cite{dirk_screw}. Another example of dynamical orbital correlation is the spin state transition in LaCoO$_3$ where the state changes gradually due to short-range orbital order~\cite{tokura_nagaosa}. A strong suppression of the Jahn-Teller-type crystal field with the lowering of structural symmetry was found in oxides such as LaVO$_3$ upon cooling~\cite{Kim_Min}. Moreover, KCuF$_3$, a material with Kugel-Khomskii-type orbital order and similar to high-T$_c$ superconducting cuprate perovskites, exhibits a large temperature range of structural fluctuations starting at the high-temperature structural transition, and extending down to the lower structural transition, which corresponds to a freezing of the octahedral rotations~\cite{lee}.

\section{\label{Conc}Conclusions}

%Our heat capacity, Raman scattering, neutron and X-ray diffraction experiments reveal an extended fluctuation regime that critically depends on $x$. . Orbital ordering is found to be reached above the structural Jahn-Teller transition with suppressed orbital fluctuations below $T_{JT}$ in $x=2.9$ and $x=2.8$. Our investigations demonstrate the stabilisation of the monoclinic $C2/c$ space-group symmetry below a characteristic temperature $T^{*}$ in Ba$_{3-x}$Sr$_{x}$Cr$_{2}$O$_{8}$. We suggest the presence of a strong orbital-lattice fluctuations over an extended temperature range, with a static behaviour only at temperatures above $T_{JT}$ and below  $T^{*}$. The Raman spectra change drastically with decreasing temperatures and the coherent part of lattice distortions develops only at lower temperatures.\par

Our heat capacity, Raman scattering, neutron and X-ray diffraction experiments reveal an extended orbit-lattice fluctuation regime that depends on Sr/Ba concentration.  Orbital ordering with suppressed orbital fluctuations is found to be reached far below the structural Jahn-Teller transition $T_{JT}$ in $x=2.9$ and $x=2.8$. Our investigations demonstrate the stabilization of the monoclinic $C2/c$ space-group symmetry below a characteristic temperature  $T^{*}$ in Ba$_{3-x}$Sr$_{x}$Cr$_{2}$O$_{8}$. %We report on the presence of a strong orbital-lattice fluctuations over an extended temperature range, with a static behavior only at temperatures above $T_{JT}$ and below  $T^{*}$. 
The Raman spectra change drastically with decreasing temperatures and the coherent part of lattice distortions develops only at lower temperatures.

%The \textit{dynamic} structural phase distortion, discussed here, is caused by strong orbit-lattice fluctuations. We have discussed different scenarios leading to a decrease in $T_{JT}$ and the suppression of the fluctuation regime upon increasing Ba content. Continuous tuning of intradimer interaction $J_0$ as well as the ratio of inter- to intradimer interaction $J'/J_0$ with $x$ could be the reason for the observed different dynamics. Change in the temperature range of the fluctuation regime with $x$ contrasts to the linear behaviour in the unit cell volumes, atomic positions and intradimer spin-spin exchange interactions.
%Depending on the value of $J'/J_0$ magnetically coupled dimers in the spin dimer network support the fast stabilization of the system, whereas large values of $J_0$ lead to individual decoupled dimers, making the system more susceptible to fluctuations.  

The \textit{dynamic} structural phase distortion discussed here is caused by strong orbital fluctuations. We have discussed different scenarios leading to a decrease in  $T_{JT}$ and the suppression of the fluctuation regime upon increasing Ba content. We conclude that a strong interdimer magnetic exchange interaction in the spin dimer network could support the fast stabilization of the orbit-lattice state, whereas large values of $J_0$ lead to individual decoupled dimers, making the system more susceptible to fluctuations.  
Therefore, small modifications in the three dimensional magnetic exchange interactions could be of more crucial importance to stabilize different structural/orbital states in spin-charge-orbital couple systems, e.g. cuprates, manganites, ferrites, etc., than small changes in unit cell volumes or atomic positions.

\begin{acknowledgments}
We thank B.~Lake for the access to the crystal growth laboratory, A.~T.~M.~Nazmul Islam for the technical assistance, and the Helmholtz Zentrum Berlin for the access to neutron beam time at the research reactor BER II.  We thank D.~Chernyshov for the discussion and measurements at Swiss-Norwegian beamline BM01 at the ESRF in Grenoble. We thank G. Thorkildsen and O. Zavorotynska for useful discussion. A.G. is supported by the Swiss National Science Foundation Grant No. 21-153659. D.W. acknowledges financial support from the Quantum- and Nano-Metrology (QUANOMET) initiative within project NL-4.
\end{acknowledgments}

\end{document}